\documentstyle[12pt]{article}
\begin{document}
\begin{titlepage}
\font\fortssbx=cmssbx10 scaled \magstep2
\hbox to \hsize{
\hskip.5in \raise.1in\hbox{\fortssbx University of Wisconsin - Madison}
\hfill$\vcenter{\hbox{\bf MADPH-95-913}
                \hbox{\bf TP-USl/95/05}
                \hbox{hep-ph/9510407}
                \hbox{October 1995}}$ }
\vspace{2cm}

\begin{center}
{\Large\bf Is there a chance to find heavy neutrinos\\
in future lepton colliders?} \\
\vspace{1.5 cm}
{\large\bf J. Gluza}\footnote{e-mail address:
gluza@us.edu.pl}
and {\large\bf M. Zra{\l}ek}\footnote{e-mail address:
zralek@us.edu.pl} \\
\vspace{ 0.5 cm}
Department of Field Theory and Particle Physics \\
Institute of Physics, University of Silesia \\
Uniwersytecka 4, PL-40-007 Katowice, Poland \\
and \\
Department of Physics, University of Wisconsin\\ Madison, WI 53706, USA\\
\vspace{2cm}

{\large\bf Abstract} \\
\end{center}
We examine two processes, the neutrino production process $e^+e^- \rightarrow
N\nu$ and the inverse
neutrinoless double-$\beta$ decay process $e^-e^- \rightarrow W^-W^-$ as
possible places for
discovering heavy neutrinos in future lepton linear colliders. The heavy
neutrino parameters
are bound from existing experimental data. We use only one important
theoretical input,
the lack of a Higgs triplet. As a consequence the neutrinos must have different
CP
parities. In such
models the
existing experimental bounds for mixing parameters still give a chance that
heavy neutrinos can be
observed in future $e^+e^-$ and $e^-e^-$ colliders.
\end{titlepage}
\vspace{0.5 cm}

The existence of heavy neutrinos is predicted by almost all
models beyond the standard one. The possibility of their
experimental discovery depends on their masses and couplings with
known leptons. There are models which predict very big masses for
heavy neutral fermions and very small couplings with known
particles. The so called see-saw models are of this type [1].
There are however other models too in which the lightness of the
known neutrinos is predicted by a symmetry argument [2,3]. In such
models the heavy neutrinos need not be extremely heavy and
the couplings are not connected with their masses. If such
models have something to do with reality, the predicted heavy
neutrinos can be potentially tested in low energy experiments.

As in the lepton sector, the standard model agrees very well with
experimental data and it is possible to find the bounds on heavy
neutrinos masses and their mixing angles. Experimental
observations like the effective number of neutrino species $N_{\nu}$,
lepton number violation processes $(\mu \rightarrow e
\gamma,\;\mu \rightarrow 3e,\;\mu \rightarrow e\; {\rm
conversion\; in \;nuclei})$ and neutrinoless double-$\beta$ decay give the most
stringent bounds on heavy neutrino parameters. The precise numerical
values of the bounds depend on the model which we consider. The
clearest situation is in the standard model with additional
right-handed neutrino singlets and we consider this model as an
example.

The aim of this paper is to give the precise values of the
cross sections for two specific processes
\begin{equation}
e^+e^- \rightarrow N\nu
\end{equation}
 and
\begin{equation}
e^-e^- \rightarrow W^-W^-
\end{equation}
at TeV energies,
taking into account present existing experimental limits on
model parameters. We think that in future colliders these processes can be a
good place where the
existence and
properties of heavy neutrinos will be tested. There are also other lepton
violation processes as e.g. $\gamma \gamma \rightarrow l^+l^+W^-W^- \;$,
$e^-\gamma
\rightarrow \nu_e l^-l^-W^+ \;(l=\mu,\tau)$ or $e^-\gamma \rightarrow
e^+W^-W^-$,
which indicate
the existence of heavy
Majorana neutrinos. But it was found [4] that these processes can be visible in
accelerators with
$\sqrt{s} \geq 4$--10~TeV over much of the range of the Next Linear Collider
(NLC) with $\sqrt{s}
\sim$ 0.5--2 TeV discussed up to now.
Although the processes (1) and (2) were considerd in the literature
[5,6,7] the situation is not clear, as different final conclusions are
predicted. The direct production process $e^+e^- \rightarrow
N\nu$ can test production and decay of heavy neutrinos with
masses up to $\sqrt{s}$ independently if they are Dirac or
Majorana particles. In the inverse neutrinoless double-$\beta$
decay process $e^-e^- \rightarrow W^-W^-$ Majorana neutrinos are
$t$-channel objects so we can hope to test them even if
their masses exceed CM energy ($M_N>\sqrt{s}$).

In the lowest order the process (1) is described
by the $W$-exchange diagram in $t$ and $u$ channels for Majorana neutrinos
(only the $t$-channel for Dirac neutrinos) and $Z$ exchange in the
$s$-channel [5,8]. The appropriate diagrams are proportional to
\begin{eqnarray}
&&K^{\ast}_{Ne}K_{\nu e}\;\;\;\;\;\;\;\;\;  (t\; {\rm and}\; u\;
\rm channels),
\nonumber \\
{\rm and} \;\;\;\;\;\;\;\;\;\; && \\
&&\sum\limits_{l=e,\mu,\tau}K_{Nl}K^{\ast}_{l\nu} \;\;\;\;
 (s\rm\;channel), \nonumber
\end{eqnarray}
where $K_{Nl}$ is the analog of Kobayashi-Maskawa mixing matrix
in the lepton sector. In the considered energy range
$\sqrt{s}>0.5$ TeV the $s$-channel exchange diagram gives only
a small contribution $(<2 \%)$ [8] so the mixing between electron and
heavy neutrino $\mid K_{Ne} \mid $ will determine the size of
the cross section $\left( \mid K_{\nu e} \mid \sim 1 \right)$.

The cross section for the process (2), described by neutrino exchange in $t$
and $u$ channels, depends on the functions [7]
\begin{eqnarray}
R_{t(u)}  &=& - \sum\limits_{\rm all\; neutrinos\;
(a)} K_{ae}^2 \frac{m_a}
{\frac{1+\beta^2}{2} \mp \beta\cos{\Theta}+\frac{m_a^2}{s}}
\end{eqnarray}
where $\beta=\sqrt{1-\frac{4M_W^2}{s}}$ and $\Theta$ is CM
scattering angle. More complicated interplay between all
elements $K_{ae}$ $(a=N,\nu)$, all neutrino masses $m_a$ and
energy $\sqrt{s}$ determines the size of $\sigma \left( e^-e^-
\rightarrow W^-W^- \right) $ [7].

What kind of information do we have from existing experimental
data? \\

(i) The sum
\begin{equation}
\sum\limits_{N(heavy)} \mid K_{Ne} \mid^2 \leq \kappa^2
\end{equation}
over heavy neutrinos is bounded.
Different values are found: $\kappa^2 < 0.015$
[9], $\kappa^2<0.0054$ [10].
If we use the last LEP result for the number of light neutrino
species, $N_{\nu}=2.991 \pm0.016$ [11], we obtain
$\kappa^2<0.0045$. \\

(ii) The lack of neutrinoless double-$\beta$ decay $(\beta\beta)_{0\nu}$ gives
the
bound for light neutrinos
\begin{equation}
\mid \sum\limits_{\nu(light)}K_{\nu e}^2m_{\nu} \mid <
\kappa^2_{light}
\end{equation}
where $\kappa^2_{light}<1.1$ eV [12] or
$\kappa^2_{light}<0.68$ eV [13]. \\

(iii) From $\left( \beta\beta \right)_{0\nu}$ it is also possible
to get the bound for heavy neutrinos $( m_N \gg 1\;$GeV)
\begin{equation}
\mid \sum\limits_{N(heavy)}K_{Ne}^2\frac{1}{m_N} \mid < \omega^2.
\end{equation}
Now, as there is a problem with estimating the role of heavy
neutrinos in $\left( \beta\beta \right)_{0\nu}$, the bounds given
by various authors differ very much:
$\omega^2<5.6\cdot10^{-4}\rm\;TeV^{-1}$~[14] or
$\omega^2<5\cdot 10^{-5}\rm\;TeV^{-1}$~[15]. \\

(iv) We know also that there are no neutrinos with $m_N>45.5$ GeV
and standard model coupings to $Z_0$ [11,16] and, if neutrinos with
masses 1~GeV $\leq m_N \leq M_Z$ exist, their coupling to
$Z_0$ should be such that [17]
\begin{equation}
Br(Z \rightarrow N\nu ) \leq 10^{-5}.
\end{equation}
There are also some general constraints which come from theory. \\

(v) First of all the mixing matrix $K$ must be unitary so
\begin{equation}
\sum\limits_{\nu(light)} \mid K_{\nu e}
\mid^2=1-\sum\limits_{N(heavy)}\mid K_{Ne} \mid^2.
\end{equation}

(vi) There are also some specific constraints connected with the
model. In gauge models the lack of Higgs triplets prevents
the production of mass terms for left-handed neutrinos. Then
the relation between light and heavy neutrinos follows [7]
\begin{equation}
\Delta_{light} \equiv \sum\limits_{\nu (light)}K_{\nu
e}^2m_{\nu}=-\sum\limits_{N(heavy)}K_{Ne}^2m_N \equiv -
\Delta_{heavy},
\end{equation}
which is crucial for our considerations.

This is all information which we use. How big could the cross
sections $\sigma(e^+e^- \rightarrow N\nu)$ and $\sigma(e^-e^-
\rightarrow W^-W^-)$ be if the couplings and masses satisfy
the constraints (5)--(10)? The answer depends on the number ($n_R$)
of heavy neutrinos and their CP parities ($\eta_{CP})$.

We would like to
clarify the point about CP parities of heavy neutrinos. From
Eqs.(10) and (6) it follows that
\begin{equation}
\mid \sum\limits_{N(heavy)}K_{Ne}^2m_N \mid < \kappa^2_{light}.
\end{equation}
As $\kappa^2_{light}$ is very small it is difficult to imagine any
model which gives so small $\mid K_{Ne} \mid $ that for
$m_N>100$ GeV the bound (11) is still satisfied. Even the
see-saw mechanism where $K_{Ne} \sim \frac{1\;GeV}{m_N}$ is not
able to give such small coupling. So the only natural explanation
is that there must be some cancellation in $\Delta_{heavy}$ and
$K_{Ne}$ are complex numbers. We do not want to study general CP
symmetry violation. We restrict ourselves to the case of CP
symmetry conservation. Then it is natural to assume that
\underline{$\eta_{CP}$ parities of heavy neutrinos are not all
equal}, as a consequence some $K_{Ne}$ are pure complex
numbers and (11) can be satisfied even for the heavy neutrinos with
$m_N \geq 100$ GeV.

First we calculated the cross section for production of heavy
neutrinos $\sigma(e^+e^- \rightarrow N\nu )$ with 1~GeV $< m_N
< M_Z$ using the bound (8). The cross section which for LEP I
is out of range of observability, $\sigma \simeq 2.5$ fb for
$m_N$=1 GeV, is larger for NLC, $\sigma \simeq 7.4$
fb for $\sqrt{s}=500$ GeV and $\sigma \simeq 8.2$ fb for
$\sqrt{s}=1$ TeV, and almost
do not depend on the neutrino mass (in the calculation we use
$K_{Ne}^2 \simeq 8\cdot 10^{-5}$ what is equivalent to the
relation (8)).

The contribution of the low mass $1\;{\rm GeV} < m_N < M_Z$ neutrinos
with small coupling (8) to the $e^-e^- \rightarrow W^-W^-$ cross
section is negligibly small ($\sigma < 2\cdot10^{-4}$ fb).

Now we restrict ourselves to larger neutrino masses $m_N> 100$ GeV.
Let us consider separately the cases with different number of
heavy neutrinos. \\

$\bullet \; n_R=1$ \\
Taking into account relations (6) and (10) the coupling $K_{Ne}$
is small
\begin{equation}
\mid K_{Ne} \mid \leq \frac{\kappa^2_{light}}{m_1}
\end{equation}
and the cross sections for both processes are very small. The
result does not depend on $\eta_{CP}$ of the heavy neutrino.\\

$\bullet\; n_R=2$ \\
In agreement with our discussion we have to assume that both
heavy neutrinos have opposite CP parities. Let us take
$\eta_{CP}(N_1)=-\eta_{CP}(N_2)=i$. If we denote
$K_{N_1e}=x_1,\;K_{N_2e}=ix_2,\;m_1=M,\;m_2=AM$ $(A>1)$ then from
relations (5)--(7) couplings and masses must satisfy the inequalities
\begin{equation}
x_1^2 \leq A\frac{\kappa^2-\delta}{A+1}+\delta \;\;\;{\rm
or}\;\;\;
x_1^2 \leq \frac{A^2\omega^2 M-\delta}{A^2-1}
\end{equation}
and
\begin{equation}
x_2^2 \leq \frac{\kappa^2-\delta}{A+1}\;\;\;{\rm or}\;\;\;
x_2^2 \leq
\frac{A}{A^2-1}(\omega^2 M-\delta ),
\end{equation}
where $\delta=\frac{\Delta_{light}}{M}$.
As for masses
$0.1\; {\rm TeV} < M < 1 {\rm TeV}, \; \kappa^2 \gg \omega^2
M$, the
second inequalities are usually stronger. The only possible way to
get large $x_1^2$ is to assume that $A \rightarrow 1$. Then the
cross section for $e^+e^- \rightarrow N\nu $ process can be
large. In Fig.~1 we depict cross sections for production of
heavy neutrinos in the $e^+e^- \rightarrow N\nu$ process as a
function of lighter neutrino mass for different values of
$A=\frac{m_2}{m_1}$ with  $\sqrt{s}=1$~TeV. There is space for
large $\sigma$ but only for very small mass differences $(m_1
\simeq m_2)$.

The  $e^-e^- \rightarrow W^-W^-$ process still remains small
and out
of `experimental interest' ($\sigma < 10^{-4}$ fb).
The functions
$R_{t(u)}$ which determine the magnitude of the $\sigma(e^-e^-
\rightarrow W^-W^-)$ prefer different masses for heavy
neutrinos $(A\gg 1)$. In the $n_R=2$ case the bound (7)
from $(\beta\beta)_{0\nu}$ has an important consequence.
Without this restriction the cross section would be
significantly greater [7]. \\

$\bullet\; n_R=3$ \\
We assume that $\eta_{CP}(N_1)=\eta_{CP}(N_2)=-\eta_{CP}(N_3)=i$.
If we denote $K_{N_1e}=x_1,\; K_{N_2e}=x_2,\; K_{N_3e}=ix_3$ and
$m_1=M,\; m_2=AM,\; m_3=BM$, then relations (5)--(7) give a set of
inequalities. We consider the more interesting case $A>B$ (if $A<B$
the mixing parameters are much smaller)
in which the following inequalities are satisfied
\begin{equation}
x_2^2 \leq -x_1^2\frac{1+B}{A+B}+\left(\kappa^2+\frac{\delta}{B}
\right) \frac{B}{A+B},
\end{equation}
\begin{equation}
x_2^2 \geq x_1^2\frac{B^2-1}{A^2-B^2}A-\left(
\omega^2M-\frac{\delta}{B^2} \right) \frac{AB^2}{A^2-B^2},
\end{equation}
and
\begin{equation}
x_2^2 \leq x_1^2 \frac{B^2-1}{A^2-B^2}A+ \left( \omega^2
M+\frac{\delta}{B^2} \right) \frac{AB^2}{A^2-B^2}.
\end{equation}
$x_3^2$ can be found from the relation
\begin{equation}
x_3^2=\frac{1}{B} \left( x_1^2+Ax_2^2-\delta \right) .
\end{equation}
{}From inequalities (15)--(17) we can find the region in the
$(x_1^2,x_2^2)$ plane of still acceptable mixing parameters. The
region (which is schematically shown in Fig.~2) depends on the
chosen values of $M,A$ and $B$.

In Fig.~3 we
depict the largest possible cross section for the $e^-e^-
\rightarrow W^-W^-$ process as a function of the lightest
neutrino mass $M$ for several values of $\sqrt{s}$. For each value
of $M$ we found the region in $(x_1^2,x_2^2)$ plane such that
values of $x_1^2,x_2^2,x_3^2$ from this region give the biggest
possible $e^-e^- \rightarrow W^-W^-$ cross section. This
situation takes place for very heavy second ($A>>1$) and
heavier third neutrino, $B \sim (2-10 )$. In Fig.3
we depict also the cross section for production of
the lightest heavy neutrino with mass $M$ in the $e^+e^-
\rightarrow N\nu $ process, taking exactly the same
mixing angle $x_1^2$ as for the $e^-e^- \rightarrow W^-W^-$
process (the curves do not represent the maximal cross section
in this case; see later in the text).  The plots presented are in some sense
model independent.
The only theoretical inputs are unitarity relation for $K$ matrix
(which is obvious) and lack of Higgs triplets (which is less
obvious and model dependent). We do not use any other
restriction as e.g.\ requirement of lack of cancellations~[4]. If we
set `the discovery limit' on the $\sigma=0.1$ fb level (which
with the year integrated luminosity $\sim 80\;\rm fb^{-1}$ [4] is
 reasonable) we can conclude that
\begin{itemize}
\item everywhere in the possible region of phase space the
production of heavy neutrinos in the $e^+e^-$ process has a
greater cross section than the lepton violating process
$e^-e^-$. It is impossible to find the place in the $(x_1^2,x_2^2)$ plane where
it is opposite.
Large values of $\sigma \left( e^+e^-
\rightarrow N\nu \right) $ makes this process a good place for
heavy neutrino searching and worth more detailed future
studies (decay of heavy neutrinos, background from other
channels [18]).
\item there are also regions of heavy neutrino masses outside
the phase space region for $e^+e^-$ where the $\Delta L=2$ process
$e^-e^-$ is still a possible place to look for heavy neutrinos. It
is a small region $1\;{\rm TeV}<M<1.1\;{\rm TeV}$ for $\sqrt{s}=1$~TeV,
$1.5{\rm\;TeV}<M<2{\rm\;TeV}$ for $\sqrt{s}=1.5$ TeV and
$2{\rm\;TeV}<M<3.1{\rm\;TeV}$ for $\sqrt{s}=2$~TeV where the cross section
$\sigma \left( e^-e^- \right) $ is still above the `discovery
limit'. There is no such place with the $\sqrt{s}=0.5$ TeV
collider. The experimental value $\kappa^2$ (see Eq.(5)) would have to
be below $\sim 0.004,\sim 0.003,\sim 0.002$ for
$\sqrt{s}=1,1.5,2$ TeV respectively to cause these regions
to vanish. Fortunately these results do not depend on the value
of $\omega^2$ (Eq.(7)) which is not well known.
\end{itemize}
If we take the other $\eta_{CP}$ parities for heavy neutrinos
our final conclusion will not change. First of all only relative
$\eta_{CP}$ are important so only one additional combination
$\eta_{CP}(N_1)=-\eta_{CP}(N_2)=-\eta_{CP}(N_3)=i$ should be
considered. This mean that the CP parity of the second neutrino
is opposite in comparison to the case which was considered
previously. The largest cross section from Fig.~3 is obtained
fora  very heavy second neutrino $(A\gg 1)$. Heavy neutrinos give
however a small contribution to the $R_{t(u)}$ function (Eq.(4)) so
it is not important if their CP parities are changed.

In Fig.~3 we do not give the experimentally
acceptable highest cross section for the $e^+e^- \rightarrow N\nu$ process. As
we mentioned before
the $\sigma \left( e^+e^- \rightarrow N\nu \right)$ depends only on
$K_{N_1e}=x_1$ mixing angle. In
the case $n_R=3$ the maximum value of $x_1^2$ is the same as in the $n_R=2$
case: $\left( x_1^2
\right)_{max} \simeq \frac{\kappa^2}{2}$. So the highest possible cross section
is the same as in Fig.~1 (continuous line). \\

$\bullet\; n_R>3$ \\
We do not obtain quantitatively new results in this case. The freedom in mixing
parameter space for
$n_R>3$, essential for our purpose, is the same as in the $n_R=3$ case. If
neutrinos have different
CP parities the relation
\begin{equation}
\sum\limits_{i=1}^{n_R} x_i^2 \leq \kappa^2
\end{equation}
determines the values of $\sigma_{max}$ and still $\left( x_1^2
\right)_{max} \simeq \frac{\kappa^2}{2}$.
The possible maximum values of cross sections are such as in the $n_R=3$ case.

In conclusion, we have found the cross sections for $e^+e^- \rightarrow N\nu$
and $e^-e^-
\rightarrow W^-W^-$ processes using the known up-to-date experimental bounds on
heavy neutrino
mixing parameters. The obtained cross sections are calculated in the standard
model with additional
right-handed neutrino singlets. The only important theoretical assumption was
that at the tree
level the left-handed neutrinos do not produce Majorana mass terms. This
had a consequences that either CP
symmetry was violated or, if it was satisfied, the CP parities of neutrinos
were
not equal. With these
theoretical assumptions we have found the `maximal possible' cross sections for
production of the heavy
neutrino process ($e^+e^- \rightarrow N\nu$) and for the inverse neutrinoless
double-$\beta$ decay
process ($e^-e^- \rightarrow W^-W^-$) in the energy range interesting for
future
lepton colliders
(0.5--2 TeV). The upper values for the cross sections were still large enough
\underline{to be
interesting from an experimental point of view}. For the $e^+e^- \rightarrow
N\nu$
process the cross
section could be as large as 275 fb for $\sqrt{s}=1$ TeV and
$M=100$ GeV. The $e^-e^-
\rightarrow W^-W^-$ process could give indirect indication for larger massive
Majorana neutrino's
existence which was not produced in $e^+e^- $ scattering. We would like to
stress once more that
what we have found are only `upper bounds' and the reality
need not be so optimistic.

\section*{Acknowledgements}
We would like to acknowledge the warm hospitality of the Institute for
Elementary
Particle Physics Research at the University of Wisconsin-Madison, where part of
this work was performed.
This work was supported in part by Polish Committee for Scientific Researches
under Grant No.~PB 659/P03/95/08, by the Curie Sk\l odowska grant MEN/NSF
93-145, by the U.S.~Department of Energy under Grant No.~DE-FG02-95ER40896 and
by the University of Wisconsin Research Committee with funds granted by the
Wisconsin Alumni Research Foundation.

\section*{References}
\frenchspacing
\newcounter{bban}
\begin{list}
{$[{\ \arabic {bban}\ }]$}{\usecounter{bban}\setlength{\rightmargin}{
\leftmargin}}
\item T. Yanagida, Prog. Theor. Phys. {\bf B135} (1978) 66; M.~Gell-Mann,
P.~Ramond and R.~Slansky, in `Supergravity', eds. P.~Nieuwenhuizen and
D.~Freedman (North-Holland, Amsterdam, 1979) p.315.
\item D. Wyler and L. Wolfenstein, Nucl. Phys. {\bf B218} (1983) 205;
R.N.~Mohapatra and J.W.F.~Valle, Phys. Rev. {\bf D34} (1986) 1642;
E.~Witten, Nucl. Phys. {\bf B268} (1986) 79;
J.~Bernabeu et al., Phys. Lett. {\bf B187} (1987) 303;
J.L.~Hewett and T.G.~Rizzo, Phys. Rep. {\bf 183} (1989) 193;
P.~Langacker and D.~London, Phys. Rev. {\bf D38} (1988) 907; E.~Nardi,
Phys. Rev. {\bf D48} (1993) 3277; D.~Tommasini, G.~Barenboim, J.~Bernabeu and
C.~Jarlskog, Nucl. Phys. {\bf B444} (1995) 451.
\item L.N.~Chang, D.~Ng and J.N.~Ng, Phys. Rev. {\bf D50} (1994) 4589.
\item G. Belanger, F. Boudjema, D. London and H. Nadeau, hep-ph/9508317
\item F. Del Aquila, E. Laermann, and P. Zerwas, Nucl. Phys. {\bf B297} (1988)
1; E.~Ma and J.~Pantaleone, Phys. Rev. {\bf D40} (1989) 2172; W.~Buchm\"uller
and C.~Greub, Nucl. Phys. {\bf B363} (1991) 349;  J.~Maalampi, K.~Mursula and
R.~Vuopionpera, Nucl. Phys. {\bf B372}, 23 (1992).
\item T.~Rizzo, Phys.Lett.{\bf B116} (1982)23; D.~London,
G.~Belanger
and J.N.~Ng, Phys.Lett. {\bf B188} (1987)155; J.~Maalampi, A.
{}~Pietila
and J.~Vuori, Nucl.Phys. {\bf B381} (1992)544; Phys. Lett. {\bf B297} (1992)
327;
C.A.~Heusch and P.~Minkowski,  Nucl. Phys. {\bf B416} (1994) 3
and `A strategy
for discovering  heavy neutrinos', preprint BUTP-95/11,
SCIPP95/07; P.~Helde, K.~Huitu, J.~Maalampi, M.~Raidal,
Nucl.Phys. {\bf B437} (1995)305;
J.~Gluza and M.~Zra\l ek, hep-ph/9502284; T.~Rizzo hep-ph/9510296.
\item J.~Gluza and M.~Zra\l ek, hep-ph/9507269.
\item J.~Gluza and M.~Zra\l ek, Phys. Rev. {\bf D48} (1993) 5093.
\item E.~Nardi, E.~Roulet and D.~Tommasini, Nucl. Phys. {\bf B386} (1992) 239;
A.~Ilakovac and A.~Pilaftsis, Nucl. Phys. {\bf B437} (1995) 491.
\item A.~Djoudi, J.~Ng and T.G.~Rizzo, hep-ph/9504210.
\item D.Schaile (LEP Collaborations) `Recent LEP Results and Status of the
Standard Model', lecture
given on the XIX Silesian School on Theoretical Physics, Szczyrk
 19--26 Sept. 1995, to appear in the Proceedings.
\item T. Bernatowicz et. al., Phys. Rev. Lett. {\bf 69} (1992) 2341.
\item A. Balysh et al. (Heidelberg-Moscow Coll.), Proc. of the International
Conference on High Energy
Physics, 20--27 July 1994, Glasgow, ed. by P.J.~Bussey and I.G.~Knowles,
vol.II, p.939.
\item J.D. Vergados, Phys. Rep. {\bf 133} (1986) 1.
\item B. Kayser, private communication.
\item  Particle Data Group, Phys. Rep. {\bf D50} (1995) 1173.
\item L3 Collaboration, O. Adriani et al., Phys. Lett. {\bf B295} (1992) 371
and {\bf B316} (1993) 427.
\item J. Gluza, D. Zeppenfeld, M.Zra\l ek, in preparation.
\end{list}
\section*{Figure Captions}
\newcounter{bean}
\begin{list}
{\bf Fig.\arabic
{bean}}{\usecounter{bean}\setlength{\rightmargin}{\leftmargin}}
\item The cross section for the $e^+e^- \rightarrow N\nu$ process as a function
of lighter heavy
neutrino mass $m_1=M$ for $\sqrt{s}=1$ TeV in the models with two heavy
neutrinos ($n_R=2$) for
different values of $A=\frac{m_2}{m_1}$ (solid line with $A=1.0001$,
`$\diamond$' line with
$A=1.004$, dots line with A=1.01 and `$\ast$' line with A=100). Only for very
small mass difference $A
\sim 1$ do existing experimental data leave the chance that the cross section
is large, e.g.\
$\sigma_{max}(M=100\;GeV) =275$~fb. If $m_2\gg m_1$ then the cross section must
be small, e.g.\ for
$A=100$, $\sigma_{max}(M=100\rm\;GeV) \simeq0.5$ fb. The solid line gives also
$\sigma_{max}(e^+e^-\rightarrow N\nu)$ for $n_R>2$ (see the text).
\item  Sketch of the region in $x_1^2-x_2^2$ plane of still experimentally
acceptable mixing
parameters. We use the following denotations (see Eqs.~(15--17) in the text)
\begin{eqnarray*}
a_1=\frac{1+B}{A+B},\;b_1=\left( k^2+\frac{\delta}{B} \right)
\frac{B}{A+B},\;a_2=A\frac{B^2-1}{A^2-B^2}&& \\
and \;\;\;\;\;\;\;\;\;\;\;\;\;\;\;\;\;\;\;\;\;\;\;\;\; && \\
 b_2=\left( \omega^2M-\frac{\delta}{B^2} \right) \frac{AB^2}{A^2-B^2},\;
b_2'=\left( \omega^2M+\frac{\delta}{B^2} \right) \frac{AB^2}{A^2-B^2}
\end{eqnarray*}
For masses $M<1$ TeV, $b_2 \sim b_2' \ll 1$ and the region is very narrow
$\left( \Delta \rightarrow 0 \right)$. The more shadowed region is the place
where the cross sections are
the largest.
\item  The cross sections for the $e^+e^- \rightarrow N\nu$ and $e^-e^-
\rightarrow W^-W^-$
processes as a function of the lightest neutrino mass $m_1=M$ for different CM
energy (the
curves denoted by F05, F10, F15 and F20 depicted the cross section for both
processes
for $\sqrt{s}=$0.5, 1, 1.5 and 2 TeV respectively).
The cross sections for the $e^-e^- \rightarrow W^-W^-$ process are chosen to be
largest. For the
$e^+e^- \rightarrow N\nu$ reaction the cross section for each of neutrino
masses is calculated using
the same parameters as for $\sigma (e^-e^- \rightarrow W^-W^-)$ and is not the
biggest one (see
the text and solid line in Fig.~2 for the maximum of $e^+e^- \rightarrow
N\nu$). The solid line parallel to the
$M$  axis  gives the predicted `discovery limit' $(\sigma=0.1\;$fb) for both
processes.
\end{list}

\newpage
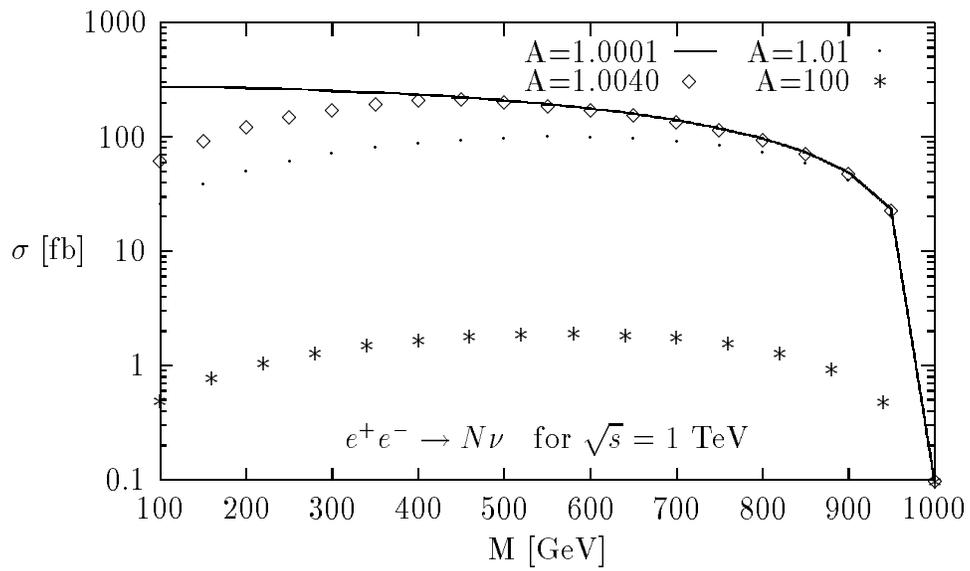
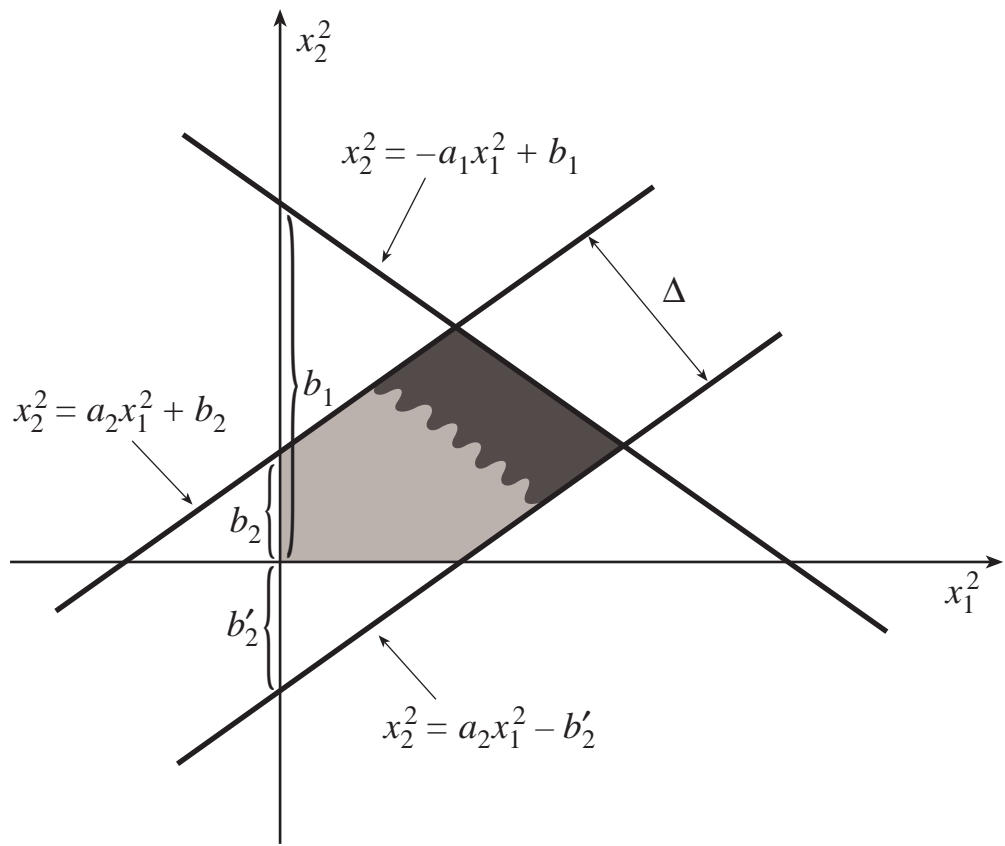
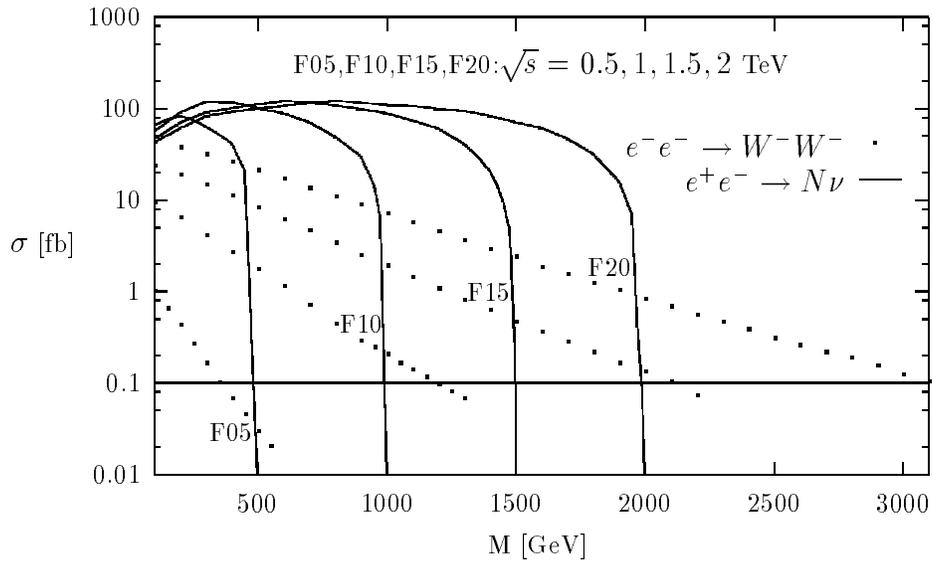
\end{document}